\documentclass[useAMS,usenatbib]{mn2e}

 \usepackage{graphicx}

\title[Fossil signatures of a merger at M31 in the Local Group]{The vast thin plane of M31 co-rotating dwarfs: an additional fossil signature of the M31 merger and of its considerable impact in the whole Local Group}
\author[Francois Hammer et al. ]{Fran\c cois Hammer$^{1}$\thanks{E-mail: francois.hammer@obspm.fr},Yanbin Yang$^{1,2}$, Sylvain Fouquet$^{1}$, Marcel S. Pawlowski$^{3}$, 
 \newauthor Pavel Kroupa$^{3}$,  Mathieu Puech$^{1}$, Hector Flores$^{1}$, Jianling Wang$^{2}$\\
$^{1}$Laboratoire GEPI, Observatoire de Paris, CNRS-UMR8111, Univ Paris Diderot, 5 place Jules Janssen, 92195 Meudon France\\
$^{2}$National Astronomical Observatoires, Chinese Academy of Sciences, 20A Datun Road, Chaoyang District, Beijing 100012, China\\
$^{3}$Argelander Institute for Astronomy, University of Bonn, Auf dem H¬ugel 71, D-53121 Bonn, Germany
}
\begin{document}

\date{Accepted 2013 March 07; Received 2013 March 06; in original form 2013 February 03}

\pagerange{\pageref{firstpage}--\pageref{lastpage}} \pubyear{2013}

\maketitle

\label{firstpage}

\begin{abstract}
The recent discovery by \cite{Ibata13} of a vast thin disk of satellites (VTDS) around M31 offers a new challenge for the understanding of the Local Group properties. This comes in addition to the unexpected proximity of the Magellanic Clouds (MCs) to the Milky Way (MW), and to another vast polar structure (VPOS), which is almost perpendicular to our Galaxy disk. We find that the VTDS plane is coinciding with several stellar, tidally-induced streams in the outskirts of M31, and, that its velocity distribution is consistent with that of the Giant Stream (GS). This is suggestive of a common physical mechanism, likely linked to merger tidal interactions, knowing that a similar argument may apply to the VPOS at the MW location. Furthermore, the VTDS is pointing towards the MW, being almost perpendicular to the MW disk, as the VPOS is. 

We compare these properties to the modelling of M31 as an ancient, gas-rich major merger, which has been successfully used to predict the M31 substructures and the GS origin. We find that without fine tuning, the induced tidal tails are lying in the VTDS plane, providing a single and common origin for many stellar streams and for the vast stellar structures surrounding both the MW and M31. The model also reproduces quite accurately positions and velocities of the VTDS dSphs. Our conjecture  leads to a novel interpretation of the Local Group past history, as a gigantic tidal tail due to the M31 ancient merger is expected to send material towards the MW, including the MCs. Such a link between M31 and the MW is expected to be quite exceptional, though it may be in qualitative agreement with the reported rareness of MW-MCs systems in nearby galaxies. 
\end{abstract}


\section{Introduction}

The Local Group baryonic content is dominated by two massive galaxies, M31 ($M_{baryon}=1.1\times 10^{11}M_{\odot}$),
 and the Milky Way (MW, $M_{baryon}=0.6\times 10^{11}M_{\odot}$), and an additional much smaller spiral galaxy, M33. Besides
this, it is populated by a plethora of dwarf galaxies (McConnachie et al., 2012) with spheroidal dwarfs (dSphs) and irregular dwarfs (dIrrs) mostly confined in
 the immediate outskirts and at large distances of the
two main galaxies, respectively \citep{Mateo98,vandenBergh06}. Such a dichotomy has been currently interpreted by the 
fact that dSph progenitors are former dIrrs having interacted with a large
galaxy, then being captured and progressively stripped of their gas \citep{Mayer07}. 

However there are now three exceptional features in the Local Group that appear uneasy to interpret, namely the
proximity of the Magellanic Clouds (MCs) near the MW and the presence of two vast structures of dSphs, including the vast polar structure
(VPOS) surrounding the MW \citep{Pawlowski12a} and the recently identified vast thin disk of satellites (VTDS) surrounding M31 \citep{Ibata13,Conn13}.
Investigation of the local volume allows us to gauge the occurrence of the MW-MCs proximity: only 0.4\% of local galaxies display such an environment \citep{Robotham12}. The occurrence of vast structures of dSphs surrounding the two main Local Group galaxies are indicative of specific geometry and motions, which are unlikely explained by motions within (large) filaments \citep{Pawlowski12b}. 

Interpreting the Local Group main features requires a {\it priori} understanding the past history of the two main galaxies.  The MW may have had an exceptionally quiet merger history \citep{Hammer07} over the past 11 billion years, which contrasts with the more turbulent history of M31 and of most spiral galaxies with similar masses \citep{Hammer09}. Two-thirds of the latter have experienced a major merger during the last 9-10 billion years, according to observations and expectations from semi-empirical $\Lambda$CDM models \citep{Puech12,Hopkins10}. In fact the M31 classical bulge and high metallicity in its outskirts both support an ancient major merger origin \citep{vandenBergh05}. The considerable number of streams in the M31 haunted halo could be the result of a major merger instead of a considerable number of minor mergers \citep[see e.g.,][hereafter H10]{Hammer10}.  This also provides a robust explanation of the Giant Stream discovered by \cite{Ibata01}: it could be made of tidal tail stars captured by the galaxy gravitational potential after the fusion time.  In fact Giant Stream stars \citep{Brown07} have ages older than 5.5 Gyr, which is difficult to reconcile with a recent minor merger \citep{Font08}. A 3$\pm$0.5:1 gas-rich merger may reproduce (H10) the M31 substructures (disk, bulge \& thick disk) as well as the Giant Stream assuming the interaction and fusion occurred 8.75$\pm$0.35 and 5.5 $\pm$0.5 Gyr ago, respectively.

In this Article, we propose that the three exceptional features (MW-MCs, VPOS and VTDS) have a common origin.  In Sect. 2 we describe the models of a merger occurring at the M31 location, which have been developed in H10 and then refined in \cite{Fouquet12}. In Sect. 3 we interpret from our modelling the mechanisms that create the exceptional features found in the Local Group underlining that perhaps, the VTDS  surrounding M31 \citep{Ibata13} has been modelled before its discovery. In Sect. 4, we investigate how such a scenario could be falsified and conclusively describes how the discovery of the VTDS may strengthen the M31 merger hypothesis and its consequences to the Local Group past history and its dwarf content.

\section{M31 modelling and consequences: the MW-MC proximity and the VPOS}

Properties of the M31 galaxy and its streams constrain fairly well a family of 3$\pm$0.5:1, gas-rich encountering models with a polar orbit (because of the 10 kpc ring), including the encountering epochs derived from stellar ages (H10). 
Because of the necessary large orbital momentum required to rebuild a giant disk as large as that of M31, a considerable amount of gas is brought near the orbital direction, to form a new disk, 5.5 billion year ago\footnote{It does not mean that all stars in the disk remnant have to be younger than 6 billion years since analyses of merger remnants indicate that older stars coming from both progenitors or formed before fusion may lie in the rebuilt disk.}. Tidal tails are included into a gigantic thick plane that is perpendicular to the orbital angular momentum, as does the M31 disk that also includes a (small) contribution from the angular momentum inherited from the main interloper \citep{Hammer10,Wang12}. In our modelling, the gigantic thick plane encompassing the tidal tails is seen edge-on from the MW as expected from the observed, almost edge-on M31 thin disk. It thus includes the MW. 
Going one step further, \cite{Fouquet12} (see their Fig. 10) identified that, within the family of M31 models, the location of the Giant Stream further limits the volume swept by one (hereafter called TT1) of the tidal tails, which is coming from material extracted at first passage, 8.5-9 billion years ago\footnote{In our modelling the Giant Stream is explained by particles coming from a tidal tail (hereafter called TT2) formed at the fusion epoch, 5.5 Gyr ago. Nevertheless its precise location constrains the location of TT1 in this family of models.}. The resulting solid angle swept by that tidal tail is found to be 5.5\% of the 4$\pi$ steradian sphere, and it still includes the MW. Such a predicted alignment by our model could be also suggested by the inclusion within 1 degree of the MW in the VTDS, if the latter is also following the orbital plane (see Sect. 3).  \cite{Fouquet12} concluded that MW dwarfs, including dSphs, could be resulting from tidal dwarf galaxies (TDGs) lying in the gigantic tidal tail (TT1) whose velocity matches the LMC proper motion. Accounting for their gravitational interactions with the MW potential, they reproduced quite well the geometrical and angular momentum properties of the MW dwarfs.

We have used a 8M particles simulations with a moderate stellar feedback in \cite{Fouquet12}, showing the formation of TDGs with locations, velocities and masses close to that of the LMC. This could lead to a simple interpretation of the LMC proximity to the MW because the LMC mass is within the observed range of TDGs \citep{Kaviraj12}. Such a configuration may appear quite unique for the Local Group, but perhaps this is linked to its specific geometry (the M31 disk orientation, the GS location and the VTDS including the MW). Because the VPOS surrounding the MW is also reproduced by such a modelling, this makes more plausible an interpretation in which many features in the Local Group are residuals caused by a gigantic encounter, involving two-thirds of its baryonic mass at the M31 location. 

 \begin{figure*}
   \begin{center}
     \begin{tabular}{c}
       \includegraphics[width=12cm]{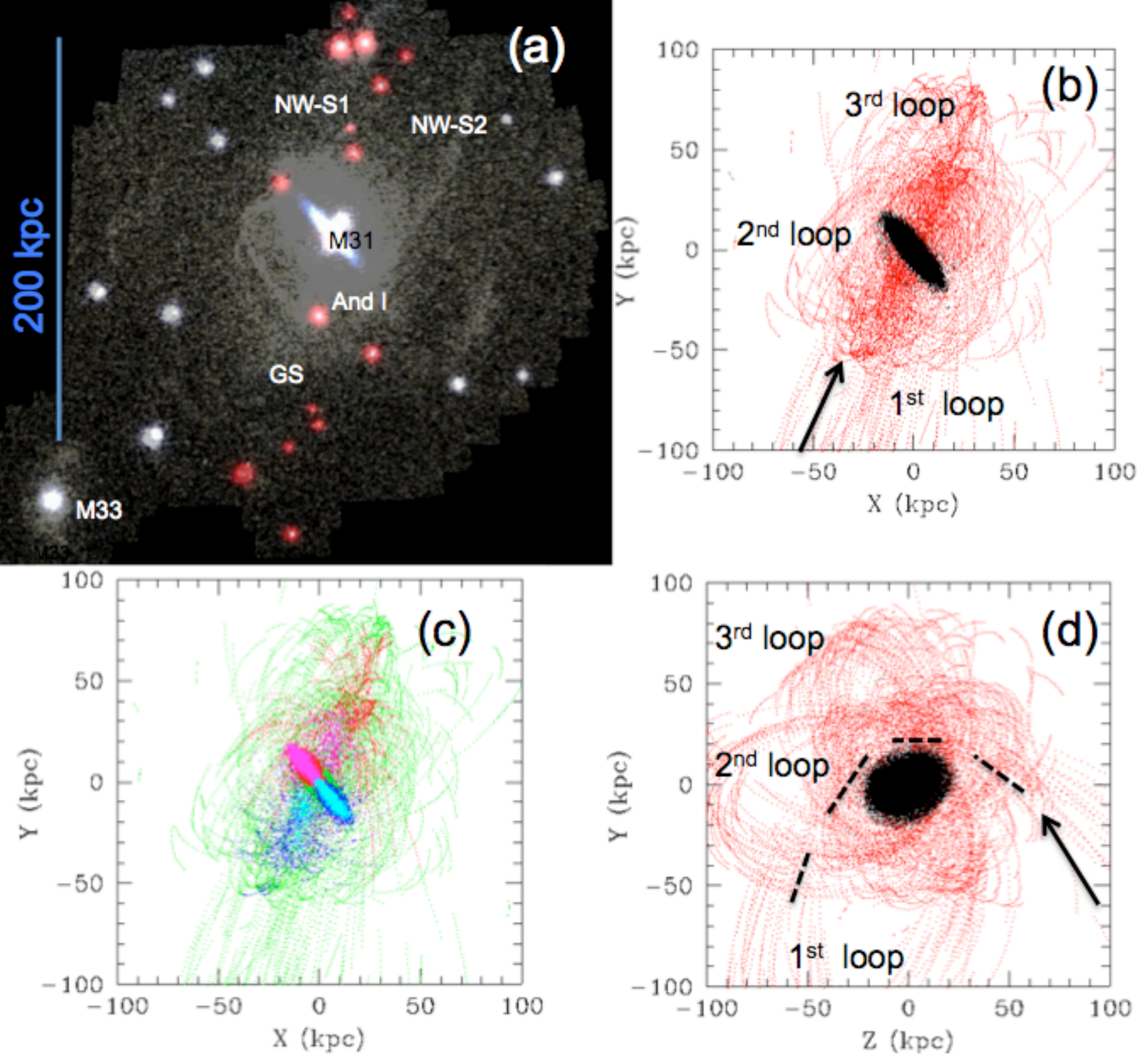}
     \end{tabular}
     \caption{{\it (a):} Capture of the VTDS (dSphs with red
colors) seen from an observer at the MW that is superimposed
to the network of stellar streams discovered by the PAndAS
 collaboration (from \citet{Ibata13}). And I lies at the GS
position and many other dSphs of the VTDS lie near the NW-S1
stream. {\it (b,c and d):}  Extracts from Fig. 8 of  H10
showing the trajectories of the star particles of a tidal
tail (TT2, see text) formed at the fusion epoch. Panel (b)
shows trajectories of particles in the observed frame. Panel
(c) is similar as panel (b),  showing heliocentric
velocities that are coded following Chemin et al. (2009),
i.e. cyan ($<$-500 km$s^{-1}$), blue (-500 to -400
km$s^{-1}$), green (-400 to -200km$s^{-1}$), red (-200 to
-100 km$s^{-1}$) and magenta ($>$ -100km$s^{-1}$). Panel (d)
is obtained after a 90 degree rotation from panel (b), for
which the plane of loops (see text) is seen almost face-on.
The black arrow in the (b,d) panels indicates the stream of
stellar particles coming from the tidal tail, which is
superimposed to the first loop in the observed frame (panel
b), as it is a generic feature for this family of M31 merger
models. In panel d, short dashed lines indicate the
approximated trajectory of particles from the tidal tail
towards the first loop.}
     \label{fig:Fig1}
   \end{center}
 \end{figure*}

 \begin{figure*}
   \begin{center}
     \begin{tabular}{c}
       \includegraphics[width=12cm]{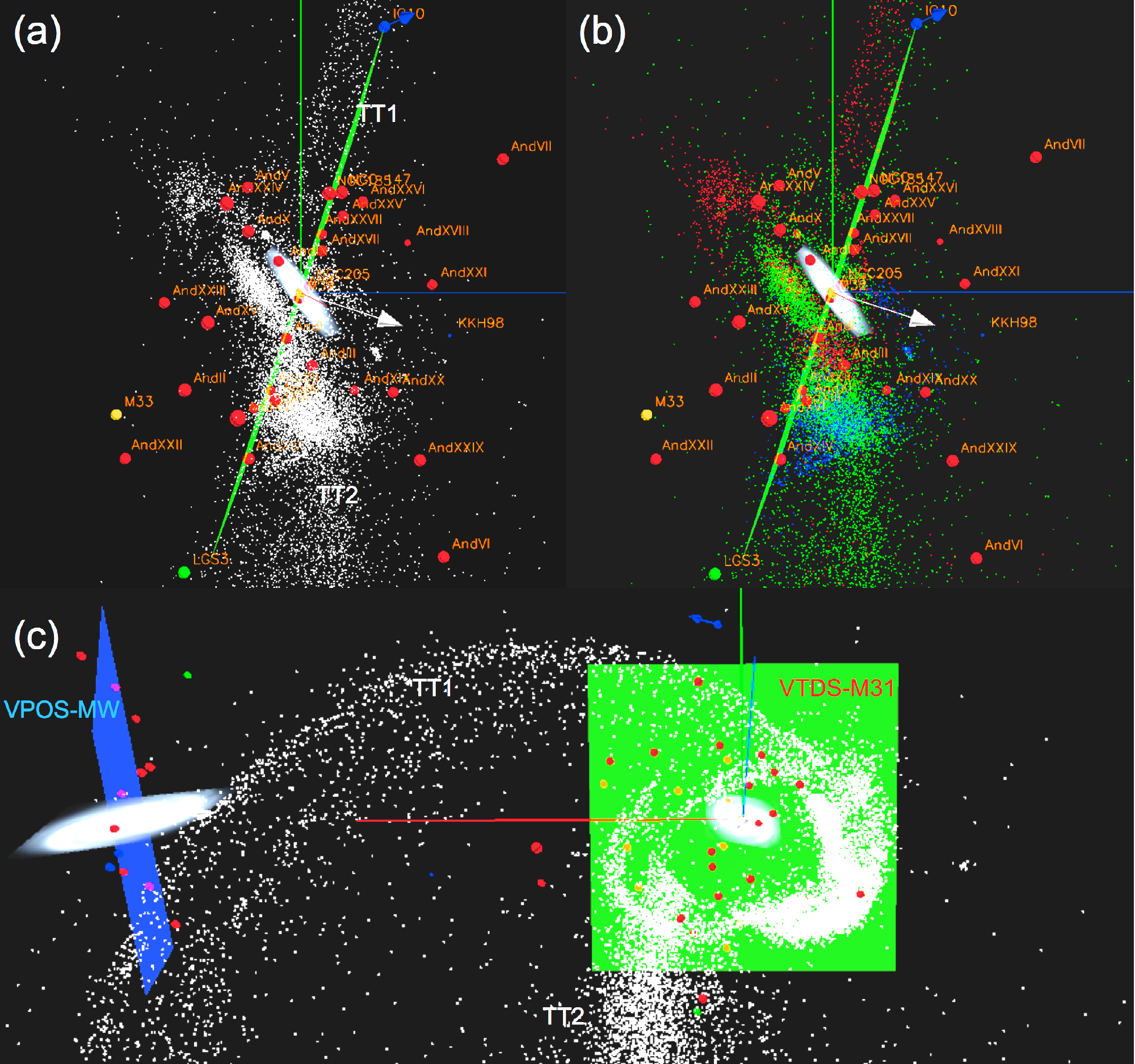}
     \end{tabular}
     \caption{{\it (a):} Overview of stellar tidal tails
(star particles shown as small white points) surrounding the
M31 remnant galaxy with the VTDS (with its dSphs shown as
red dots) being seen almost edge-on as in \citet{Ibata13}'s
Figure 3. Other dSphs are shown with white dots and dIrrs
and dTrans in blue and green dots, respectively. To allow a
clear visibility of tidal structures, all simulated
particles (from the 8M particles simulation of
\citet{Fouquet12}) within r$<$ 120kpc from the M31 center
have been removed. The simulated GS, thought too extended to
the west, almost coincides with the southern part of the
VTDS (green array) and is caused by particles coming from
TT2. The northern part of the VTDS is almost coincident with
TT1 that is a tidal tail formed at the first passage.  {\it
(b):}  Same as (a) but with heliocentric velocities coded
(with an enhanced contrast for a better view) with blue (v
$<$ -350 km$s^{-1}$), green (-350 to -250km$s^{-1}$), and
red (c $>$-250 km$s^{-1}$). {\it (c)}: Same as (a) with the
VTDS (green array) seen face-on and using a slight
de-zooming to show the MW and the VPOS (blue array, based on
the 11 classical MW dwarfs from \citet{Fouquet12}). The VTDS
is aligned within 1 degree to the M31-MW axis (red
horizontal line).}
     \label{fig:Fig2}
   \end{center}
 \end{figure*}

 \begin{figure*}
   \begin{center}
     \begin{tabular}{c}
       \includegraphics[width=5.54cm]{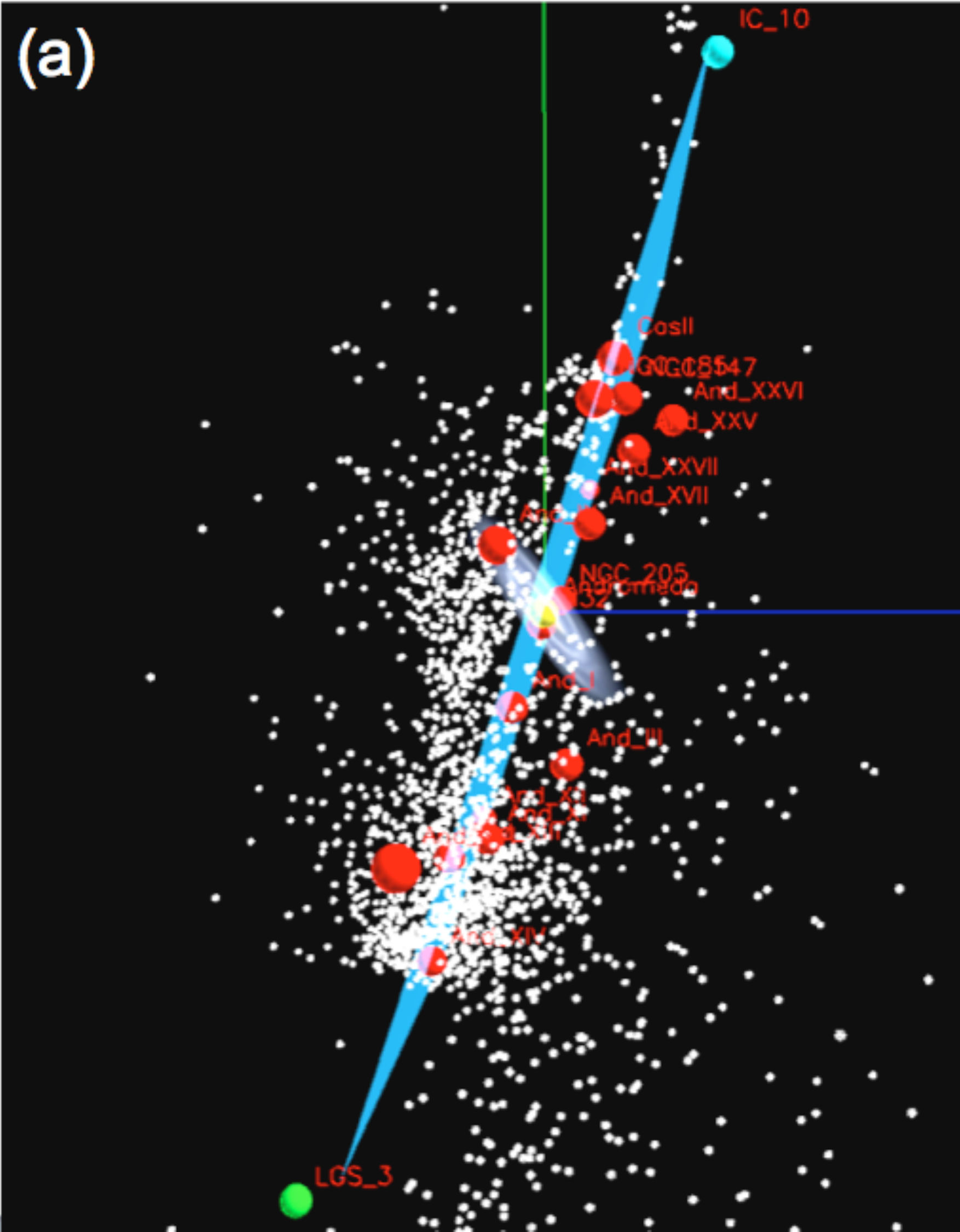}
       \includegraphics[width=8.4cm]{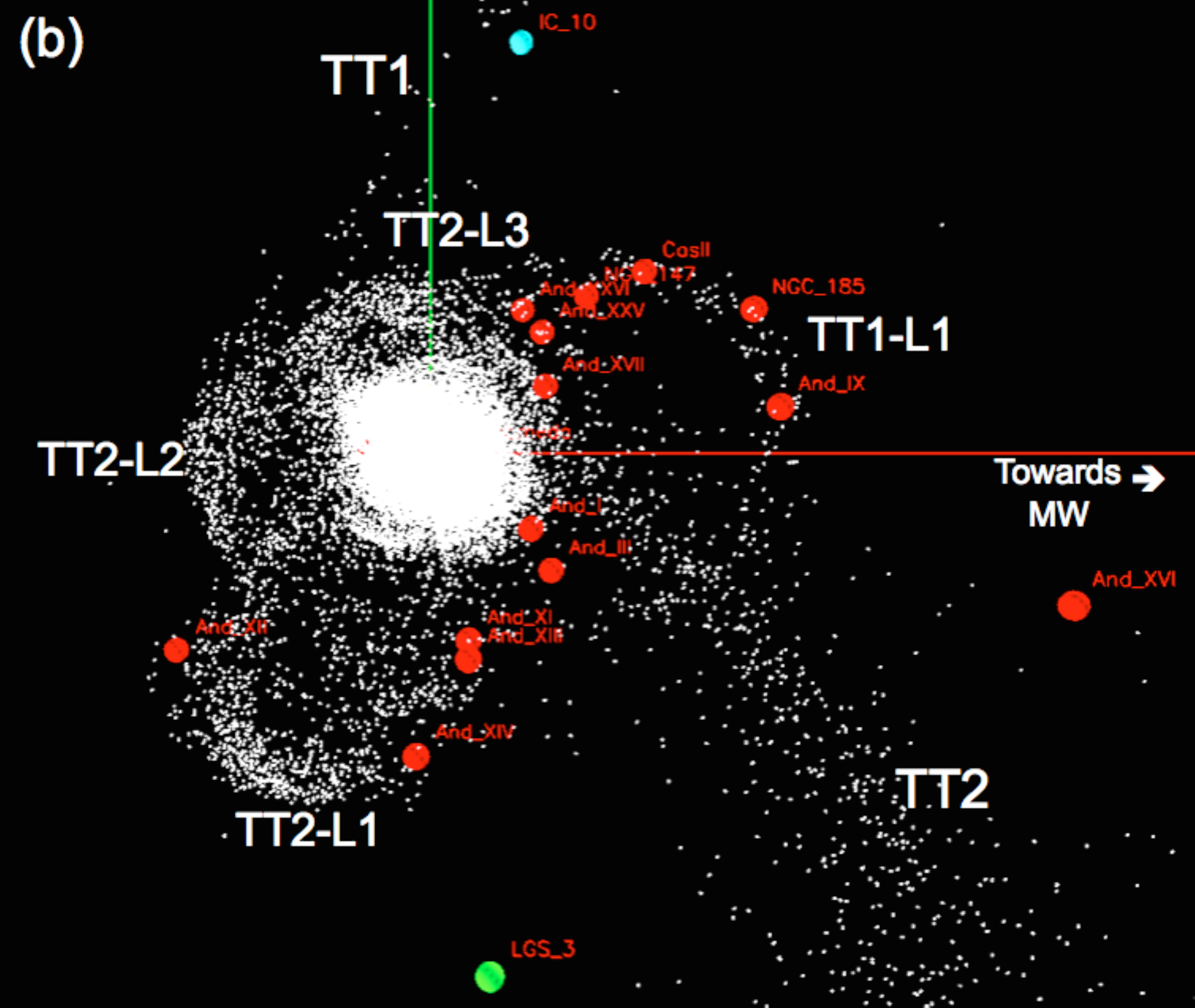}
     \end{tabular}
     \caption{{\it (a):} Similar than Figure \ref{fig:Fig2}
(a) for another model with different mass ratio (3:1 instead
of 3.5:1), which shows a better agreement between the tidal
tail systems (including loops) and the VTDS plane (blue
array).  The improvement is mainly due to the fine tuning of
the main progenitor inclination by ~20 degrees. This tuning
suffices to change the plane of tidal tail systems with
respect to the orientation of the final M31 disk plane. It
also shows that the VTDS extent could be matched by stellar
particles of the model, while IC10 is lying within TT1.{\it
(b):} Same as (a) but rotated (by $\sim$90 degrees) to have
the VTDS seen exactly face-on, with loops labelled by their
tidal tail origin (TT1 or TT2) and their loop order (L1 to
L3).  Only galaxies near the VTDS plane are shown to
illustrate at best whether or not they are matching the loop
system. In panel (b) we have chosen to remove particles
within r$<$ 20kpc from the M31 center for a better
visibility of the loop system.}
     \label{fig:Fig3}
   \end{center}
 \end{figure*}

\section{The Vast Thin Disk surrounding M31}
\subsection{A predicted feature before its discovery? }

The VTDS is a recent discovery of a thin structure including half of the M31 dSphs \citep{Ibata13}. It is even more significant \citep[0.99998 significance, see also][]{Conn13} than the VPOS, and both structures show evidence for co-rotation with an axis well offset from that of the host galaxy disk. \cite{Ibata13} discussed the possible accretion or in-situ formation scenarios, finding them not very  convincing  because of the considerable thinness of this vast structure (perpendicular scatter of 14 kpc for a $\sim$ 400 kpc diameter).

The VTDS angular momentum could be an ancient and persistent fossil of a gigantic event in the M31 past history if it is as old as its constituents, the old stellar populated dSphs. Furthermore the VTDS, the Giant Stream (GS) and the NW Streams \citep[NW-S, ][]{Lewis12} share approximately the same position angles (PAs), since And I is included in both VTDS and the Giant Stream and many VTDS dSphs are superimposed near the NW-S1 \cite[see Figure \ref{fig:Fig1}-a and also ][]{Conn13}. Moreover the VTDS dSphs in the Southern part of M31 have heliocentric velocities that are included within the broad range (-550 to -300km$s^{-1}$) of GS stars \citep{Ibata04}, i.e. approaching us faster than M31. This suggests an ancient and common origin for all these structures, leading us to examine a tidal origin for the VTDS. 

Figure \ref{fig:Fig1}(b,c,d) also illustrates the trajectories of stellar particles of a tidal tail formed at the second passage (hereafter called TT2), near the fusion time. In H10 the goal was to figure out the formation of the GS. As such, these trajectories are a generic property for all the family of M31 merger models (see their Table 3). Tidal tail particles captured by the potential of M31 have their trajectories inscribed into a series of loops (see Figure \ref{fig:Fig1} and Fig. 8 of H10) due to the absence of dynamical friction \citep[see a complete description of the loop mechanism in][]{Wang12}. The overall loop system is inserted into a plane that is relatively thin and seen almost edge-on in the observed frame, as it is illustrated in Figure \ref{fig:Fig1} (see a and b panels). Thus VTDS, GS, NW-S and the prediction of H10  share similar PA and velocity distributions, supporting our claim that perhaps the VTDS was roughly modelled before its discovery. 

However this might be only a part of the explanation, because of the large extent of the VTDS to radii of at least 200 kpc: the 1st loop is sufficiently extended to the SW side of M31 while star particles hardly catch the full extent of the VTDS (Figure \ref{fig:Fig1}) in the NW. There could be two explanations for this. Either it requires a more energetic collision that would expand the loop extent or some of the dSphs in the NW are related to another mechanism.  Figure \ref{fig:Fig2}(a) shows  the other tidal tail (TT1) formed at the first passage of the M31 merger, 8.5 to 9 billion years ago, which almost coincides with the northern extent of the VTDS. Moreover, Figure \ref{fig:Fig2}(b) indicates that particles associated to TT1 are redshifted while those from TT2 are blueshifted following precisely the observed motions of dSphs in the VTDS. We have investigated whether another plane of stellar particles could have been formed from TT1, that could be parallel to the loop plane presented in Figure \ref{fig:Fig1}. 

If the NW dwarfs are associated with TT1, the ancient tail, while the southern dwarfs are associated with TT2, the fusion tail, it becomes unclear how all may lie in the same thin plane. \cite{Conn13} suggested that the VTDS could have been associated to a 5 Gyr old event on the basis of expected orbital time, i.e., favouring an event associated to the fusion time, and then linked to TT2. Figure \ref{fig:Fig2} also reveals that TT2 includes much more stellar particles in the immediate outskirts of M31 than TT1. Moreover, during 5.5 Gyr (since the fusion time) particles have had enough time to reach high order loops as evidenced by Figure \ref{fig:Fig1} (b), a result similar to that found by \cite{Wang12} to explain the remarkable loop system in the outskirts of NGC5907. The Giant Stream is also supporting that TT2 is mostly responsible of the VTDS dSphs as it is associated in our modelling to the superposition of TT2 and its first loop. It is the more prominent reservoir of stars in the M31 outskirts while TT1 can be only associated to the extremely faint NW-Stream 1. 

\subsection{Most VTDS dSphs are distributed along the merger induced loops}

Figure \ref{fig:Fig3} provides a first glimpse for an improvement of our modelling of the VTDS. Here the merger is more energetic than that shown in Figure \ref{fig:Fig2} because of a more massive secondary. It allows stellar particles to be better aligned with the VTDS as well as to reach the full extent of the VTDS. Figure \ref{fig:Fig3}(b) reveals the associated loop planes. TT2 loop plane is very similar to that in Figure \ref{fig:Fig1}(d), evidencing that stellar particles have enough time to reach the third loop. TT1 is associated to a single loop that is less prominent, all these properties being shared by the modelling of NGC5907 loops \citep{Wang12}. From Figure \ref{fig:Fig3}(b), it seems that the VTDS dSphs are easy to associate within the loop system, except And XVI and XXVII, which are too far away to be in the loop system\footnote{And XXVII is possibly an interloper due to its counter rotating motion \citep{Ibata13}.}. Namely, we have verified from their 3D locations that: And I, III, XI, XII, XIII, and XIV lie within the TT2 first loop, as And XVII, XXV and XXVI  do with the TT2 third loop. NGC 147, 185, CasII (AndXXX) and And IX are likely associated to the TT1 first loop. One may wonder why dSphs are apparently absent from the TT2 second loop. Maybe this is because TDGs currently lie at discrete locations along a tidal tail giving it a so-called string of beads appearance \citep{Wetzstein2007,Bournaud10,Fouquet12}. Perhaps this explain why most of VTDS dSphs are observed on M31 halo side nearest the MW. We have also verified that dSph velocities are in a roughly good agreement with expectations from the loop velocities (see also Figure \ref{fig:Fig1}-c) with the noticeable exception of And XIII. It seems that most VTDS dSphs (9 among 13) are related to TT2 and that, in our modelling, the VTDS implies a good alignment between TT1 and TT2 within the (hyper)plane defined by the orbital angular momentum.



\section{Discussion and Conclusion}

 \subsection{Discussing possible falsifications}
The proposed scenario can at the same time account for the M31 Giant Stream, the VTDS around M31 \citep{Ibata13}, the fact that it points towards the MW, the vast polar structure around the MW \citep{Pawlowski12a}, the fact that the two vast structures are also rotating, and finally, the proximity of MCs to the MW. Of course implications are so vast that it would be useful to search for evidence which falsifies this scenario, especially given the impressive knowledge of the Local Group content. If not passing this step, our scenario could be considered as an interesting ballistic exercise. This is further complicated because of the enormous amount of parameters to investigate and also because hydrodynamical models \citep[GADGET2, ][]{Springel05} require significant number of particles to interpret faint stellar halo features. It might be also argued that we have no proof that other kinds of models could not fit the M31 merger. However the predicted loop system is particularly consistent with dSph locations and velocities, possibly supporting the family of models investigated in this paper.

 \subsubsection{dSph dark matter content}

If TDGs are progenitors of many dSph in both MW and M31 outskirts, it leads to an absence of dark-matter (DM) in galaxies that are being thought to be the most DM-dominated systems.
 Clearly the above ballistic exercise has to be discarded if the dark matter (DM) content of dSphs is large \citep{Strigari08} as inferred from their large velocity dispersions \citep{Walker09}. Perhaps the DM content of dSphs requires some further investigations:  dSphs could be alternatively the outcome of TDGs, which are gas-stripped when entering the halo of a large disk galaxy such as the MW \citep{Kroupa97, Casas12, Pawlowski11}. We are indeed investigating a similar interpretation but with a tidal tail coming from M31 and currently reaching the MW \citep{Yang13}. We find that simulated DM-free TDGs are quite fragile, and, helped by the large eccentricity of their orbit (due to the M31 motion towards us), are almost destroyed during a single passage, providing a fair reproduction of most dSph properties \citep{Yang13}. Most properties currently associated to DM are predicted, including large apparent M/L as they are calculated by \cite{Walker07,Walker09}, and flat radial distribution of their velocity dispersions.
  
Besides this, VTDS, M31 dSphs may have had enough time (5.5 billion years for TT2) for their gas content to be exhausted through tidal interactions with M31 \citep{Kroupa97, Casas12, Yang13}. In fact, the particle motion along the loops is simply that of test particles in a static central potential, which is a rosette \citep{Binney87,Wang12}, following trajectories within elliptical loops (see Figures \ref{fig:Fig1}-d and \ref{fig:Fig3}-b). The small pericenter at the first approach (see dashed lines in Figure \ref{fig:Fig1}-d) likely ensures an efficient tidal stripping. 

 \subsubsection{The LMC as a TDG remnant}

Such an association is indeed non trivial, though there are observed TDGs with masses equal or larger to that of the LMC \citep{Kaviraj12}. The LMC dynamical mass according to \cite{van der Marel09} is only a factor 4 larger than its baryonic mass. \cite{van der Marel09} also reviewed the difficulty in establishing $V_{rot}$ that varies significantly with the tested stellar or gas component. The LMC disk seems to be not circular and fairly affected by a bar, while $V/\sigma$=3 for carbon-stars is indicative of a thick disk. Further tests based on full 3D velocity field are essential to show whether or not the kinematics of  Magellanic dIrrs are consistent with those of TDGs (either simulated or observed).
On the other hand, TDGs (observed or simulated) show low $V/\sigma$ values, with gas extending much beyond the optical extent similarly to Magellanic dIrrs. Applying a similar method to retrieve TDG rotational curves as has been done for dIrrs, it appears that both populations lie on the baryonic Tully Fisher relation \citep{Gentile07}, providing another support for a common origin \citep{Kroupa12}.

 \subsubsection{The Mass-Metallicity relation}

A third possible falsification of the present model could come from the luminosity-metallicity relation. First, it has to be pointed out that TDGs in the first tidal tail (TT1) would have travelled for about 8.5-9 Gyr before arriving at the MW location, while those assumed to be progenitors of the VTDS dSphs may have spent several Gyr within TT2. Thus the context to be considered is the evolution of a dwarf within a gas-rich environment (the tidal tail) that may favour star formation, feedback as well as collisions between small galaxies, i.e., conditions that are similar to those invoked to explain the origin of the mass-metallicity relation (see e.g., the introduction of \citet{Foster12}). Indeed, a theoretical study by \cite{Recchi07} indicates that TDGs may fall on the mass-metallicity relation.
The large variety of star formation histories found in dSphs may match expectations if they are TDGs remnants. By the way, a part of the stellar population has to be old and with low metallicity, 
since it has been extracted from the secondary interloper, a galaxy similar to the faintest galaxies at z$\sim$ 1.5, at redshifts where faint galaxies should be metal poorer than
present-day galaxies in the same mass range \citep{Hammer09,Rodrigues12}. Even very low metallicities can be accommodated within our scenario, either from early, metal poor stars in the progenitor or from the outer part of the secondary gaseous outskirts, that could be almost primordial at such early epochs.

 \subsection{Revisiting the Local Group past history and its dwarf spheroidal content}
 
Figure \ref{fig:Fig1} and Figure \ref{fig:Fig2} evidence the peculiarity of the geometry of the Local Group main galaxies and their associated vast structures:
\begin{enumerate} 
 \item the VTDS is almost aligned with the GS PA sharing its velocity distribution, and it is also aligned with the NW Stream 1 (NW-S1);
 \item the M31 disk is almost seen edge-on from the MW;
 \item the VTDS is aligned within one degree with the M31-MW (\cite{Ibata13}, see also Figure \ref{fig:Fig2}-c);
 \item both VTDS and VPOS are perpendicular within a few degrees to the MW disk, see Figure \ref{fig:Fig2}-c.
 \end{enumerate} 
 
The link between the M31 tidal features and the VTDS (point i, above) as well as the MW tidal features and the VPOS \citep{Pawlowski12a} is suggestive of a tidal origin for both vast structures. Furthermore, the three geometrical alignments (points ii, iii and iv)  are suggestive of a common origin for both the VTDS and VPOS. If correct, this may lead to a single merger origin for these vast, fossil structures in the Local Group, which include a significant part of the dSphs that surround the two main galaxies. In fact, this could be related to the \cite{Ibata13}'s conclusion, quoting: "An alternative possibility is that gas was accreted preferentially onto dark matter sub-halos that were already orbiting in this particular plane, but then the origin of the plane of sub-haloes would still require explanation". Such an origin is precisely provided by the orbital plane of an ancient merger at the M31 location, though we assume here that dSphs are not sub-haloes but are descendants of TDGs. Gas-rich mergers are producing gas-rich tidal tails from which material is re-accreted to the host galaxy in a particular plane, here the VTDS. Because it is pointing within 1 degree to the MW \cite{Ibata13}, it may provide as well an explanation of the VPOS as it is described in \cite{Fouquet12}\footnote{Note that the angle between the VTDS and the VPOS is 51 degrees that is consistent with trajectories of the MW dwarfs within the VPOS as found by \cite{Fouquet12}.}.

Figure \ref{fig:Fig2} evidences that the VTDS and the tidal tails generated by the M31 merger model are lying in the same gigantic plane that also includes the MW, perhaps revealing a quite surprising, alternative past history for the Local Group. Nine billion years ago, the latter would have been made of three main galaxies, one being the MW, the two other being interacting with an orbital motion in the direction of the MW, before they finally merge to eventually form the present-day M31 galaxy. The induced tidal tail at the first passage would reach the MW at relatively recent epochs, with a large velocity consistent with that of the LMC \citep{Kalli09,Yang10,Fouquet12}. It could be argued that the chance for such an occurrence is particularly small. This is indeed correct and would naturally explain why MW-MC systems are so rarely found in the Local Universe. Such an occurrence is found to be only 0.4\% according to \cite{Robotham12}, and, interestingly the rare MW-MCs analogues are found in double galaxy systems such as the Local Group. 

Here, we interpret the exceptional vast planes in the Local Group as being the fossils of an ancient process occurring in the past history of the Local Group, namely an ancient merger at the M31 location. Our conjecture is perhaps the only way to explain them through a common process together with the MW-MCs proximity. It does not require a revision of the galaxy formation theory, as it belongs to the framework of the hierarchical scenario \citep{White78} and it is in agreement with expectations from observations of distant galaxies \citep{Hammer05,Hammer09,Puech12}. In the modelling we have adopted a 20\% fraction of baryons in relative agreement with cosmological parameters, while other ratios may be investigated, without evidence that it can affect the geometrical arguments described in this paper\footnote{The ballistic solutions presented here should not be radically different for Milgromian dynamics since they can be approximated by Newtonian dynamics with a phantom dark matter halo.}.

We are not claiming that we have succeeded to model the full history of M31 and of the Local Group, because many refinements are needed.  The discovery of the VTDS by \cite{Ibata13} gives us 
an extraordinary opportunity for a fine tuning of our model. For instance, in Figure \ref{fig:Fig2}(a) the simulated GS is offset to the west and is possibly too wide. We need to match more accurately a VTDS pointing to the MW and the locations of dSphs in the loop systems. Extremely large number of particles is also required to form realistic TDGs within a large mass range. More precision is needed and predictions of tidal tail occurrences, locations and angular momentum (Athanassoula, 2013, in preparation) would be extremely valuable for such a gigantic effort. Understanding whether or not some dIrrs can be associated to TDGs is also a challenge to assess or disprove our conjecture. 

Nevertheless, our modelled conjecture provides a reasonable mechanism for understanding together most puzzling features in the Local Group. If correct it would affect the $\Lambda$CDM theory as a significant part of dSphs would be devoid of dark matter through a process that reproduces most of their observed properties \citep{Yang13,Dabringhausen13}. It would severely strengthen the existing problem of missing satellites \citep[see][and references therein]{Boylan-Kolchin11,Kroupa12}, that could be better rephrased as "a significant excess of small halos" predicted by the current theory.

\section*{Acknowledgments}
We are very grateful to our referee, Brent Tully, whose comments have strongly improved the current version of this paper.
This work has been supported by the ChinaÐFrance International
Associated Laboratory ÒOriginsÓ supported by the Chinese Academy of Sciences, the National Astronomical Observatory of China, the Centre National de la Recherche Scientifique and the Observatoire de Paris. Part of the simulations have been carried out
at the High Performance Computing Center at National Astronomical
Observatories, Chinese Academy of Sciences, as well
as at the Computing Center at Paris Observatory.
The three-dimensional visualization was conducted with the S2PLOT programming library for representing the Local Group galaxies in 3D.

\bsp

\label{lastpage}

\end{document}